\documentclass[a4paper,11pt]{article}
\usepackage[a4paper, top=2.7cm, bottom=2.7cm, left=2.7cm, right=2.7cm]{geometry}
\usepackage[T1]{fontenc}
\usepackage[utf8]{inputenc}
\usepackage{lmodern}
\usepackage[round,authoryear]{natbib}
\usepackage{graphicx}
\usepackage{float}
\usepackage{lineno}
\usepackage{nameref}
\usepackage{afterpage}
\usepackage{setspace}
\doublespace
\usepackage{fancyhdr}
\title{Experiences with efficient methodologies for teaching computer programming to geoscientists}
\author{Christian T. Jacobs, Gerard J. Gorman, Huw E. Rees, Lorraine Craig}
\date{\small \textbf{Affiliation for authors \#1, \#2 and \#4:} Department of Earth Science and Engineering, South Kensington Campus, Imperial College London, London, SW7 2AZ\\\textbf{Affiliation for author \#3:} Educational Development Unit, South Kensington Campus, Imperial College London, London, SW7 2AZ\\\textbf{Corresponding author email address}: \texttt{c.jacobs10@imperial.ac.uk}\\\textbf{Short title:} Efficient methodologies for teaching programming to geoscientists\\\textbf{Paper type:} Curriculum \& Instruction (Instructional Approaches)
\newline\\\textbf{Manuscript accepted for publication}\\\textbf{in the Journal of Geoscience Education on 9 June 2016}}

\begin{document}

\maketitle
\thispagestyle{empty}
\setlength{\parskip}{0.25cm}
\setlength{\parindent}{0cm}

\begin{abstract}
% MUST BE LESS THAN 250 WORDS
Computer programming was once thought of as a skill required only by
professional software developers. But today, given the ubiquitous nature of
computation and data science it is quickly becoming necessary for all
scientists and engineers to have at least a basic knowledge of how to program.
Teaching how to program, particularly to those students with little or no
computing background, is well-known to be a difficult task. However, there is
also a wealth of evidence-based teaching practices for teaching programming
skills which can be applied to greatly improve learning outcomes and the
student experience. Adopting these practices naturally gives rise to greater
learning efficiency - this is critical if programming is to be integrated
into an already busy geoscience curriculum. This paper considers an undergraduate computer programming course, run during
the last 5 years in the Department of Earth Science and Engineering at Imperial College London. The teaching methodologies that
were used each year are discussed alongside the challenges that were
encountered, and how the methodologies affected student performance. Anonymised
student marks and feedback are used to highlight this, and also how the
adjustments made to the course eventually resulted in a highly effective
learning environment.
\end{abstract}

\section*{Introduction}\label{sect:introduction}
Computer programming is increasingly becoming an essential skill for
geoscientists as the world becomes more digitalised. We might commonly think of
computer programming as an activity carried out by a relatively small number of
domain specialists developing complex application software, perhaps in
collaboration with computer scientists. However, geoscientists and engineers
are faced with numerous day-to-day tasks such as manipulating datasets (e.g.
standardising, reformatting or filtering), statistical analysis, plotting, or
automating repetitive tasks such as rerunning the same program with many
different data inputs and gathering the results for analysis. Even in the case
of a software user, where no active programming is required, a basic knowledge
of computer programming and experience in debugging can be critical when
troubleshooting third party software. This is because programming skills
provide the user with a conceptual model for understanding what might be going
wrong, systematically characterise the problem, and then either modify their
workflow or constructively engage the software developers to resolve the
problem. Enhanced computing power is enabling simulation and data inversion to
play a greater role in discovery and prediction in geoscience. Arguably we are
rapidly approaching a point where innovations will predominately come from
those who are able to translate an idea into an algorithm, and then into
computer code.

It has long been recognised that teaching basic programming skills to
novices is difficult
\citep{robins2003learning}. \cite{winslow1996programming} suggests
that it takes about 10 years of experience to turn a novice into an
expert programmer. Worryingly, these conclusions were mostly drawn from teaching
computer science students where computing dominates the
curriculum. Therefore, it follows that careful consideration needs to
be given to the design of an introductory programming course when it
only forms a small part of a non-computing curriculum. There are also
motivational issues due in part to the subject being largely
associated with the field of computer science, and geoscientists are
therefore often surprised to see it as part of their curriculum. This
can lead to the opinion that the subject is not worth pursuing, or the
worry that they do not have the background or potential to do well in
the subject. Even in the case where the student does have a strong
computing background, learning how to program, much like learning to
swim or how to ride a bike, requires a great deal of practice. The
learning experience is likely going to be completely different to what
the student is used to. Not only is there a large amount of unfamiliar
material/knowledge to understand, the student must also adapt to
different methods of content delivery and a highly practical learning
methodology. Furthermore, the mind has to be trained to think `like a
computer' (i.e. to follow a series of steps in a logical way/as a
`process').

In light of these challenges, a considerable amount of research has
gone into developing effective strategies for teaching a course in
introductory programming; an excellent review is given by
\cite{Pears2007}. A key consideration in a geoscience context (or
indeed any course outside a dedicated computer science degree) is that
introductory programming is not being taught as part of a wider
computer science curriculum, but instead has to fit in within an
already full geoscience curriculum. One specific concern is that a
single introductory programming course is unlikely to enable the
students to take these skills and reapply them to a different
problem-solving context from where they were presented
\citep{palumbo1990programming}. Therefore, we also need to consider
where else in the curriculum there are opportunities to use and extend
students' programming skills and experience.

The choice of a first programming language also has a significant
impact on learning. While the top 3 popular programming languages have
been consistently C, Java and C++ for many years \citep{TIOBE_2015}, they are
generally not thought to be good choices as a first programming
language \citep{mody1991c, churcher1998java, biddle1998java,
  clark1998java, close2000cs1}. A big part of the problem with
starting with such programming languages is that they are too low
level, and the high cognitive load associated with the syntax
\citep{StefikSiebert_2013} has little to do with learning to think
algorithmically and writing structured programs
\citep{Pears2007}. Interestingly \cite{Pears2007} also point out that
similar learning issues related to excessive cognitive load arise when
using professional Integrated Development Environments (IDEs) in
introductory programming due to the effort that must be invested to
become a proficient user. For this reason high level languages such as
Python are a popular choice because of their much simpler,
higher-level syntax \citep{Donaldson_2003, Fangohr_2004, Lin_2012}.

\cite{boszormenyi1998java} points out that a first programming language should
not be decided in isolation but in the context of the entire curriculum. There
are many opportunities to reinforce learning if the same language can be used
as a tool in other lecture courses and project work. However, this relies upon
a common denominator being agreed among the teaching staff.  While many
computer languages (e.g. C/C++, FORTRAN, Java) are used by research and
academic staff, Python has become very popular as it is both simple and
powerful. A lot of international effort has gone into providing Python
interfaces to popular geoscience packages to the extent that Python can be used
seamlessly across laptops, supercomputers and Cloud platforms. Examples
include: GRASS GIS \citep{GRASS_GIS_software},
ArcGIS \citep{ESRI_2015}, and QGIS for GIS and
geomorphology \citep{QGIS_software}; PyLith for modelling crustal deformation
\citep{aagaard2007pylith}; ObsPy for processing seismology data
\citep{beyreuther2010obspy}; and Firedrake for geophysical fluid dynamics
\citep{Rathgeber2015, Jacobs2015}. This provides an environment where students
feel motivated from day one that they are learning a language that can be
directly applied professionally, while still benefiting from a simple syntax
with relatively low cognitive overhead so that they can focus on learning to
think algorithmically.

In 2010 we began to develop an 8-week introductory programming course with 24 contact hours for undergraduate students, majoring in geoscience, in the Department of Earth Science and Engineering at Imperial College London\footnote{Originally this was a second year course. In 2012 we had to transition the course from being a `second year only' course to being a `first year only' course; this was accomplished by teaching both first year (i.e. the 2012 intake) and second year (i.e. the 2011 intake) students in the same course that year.}. The objective of the course was to teach fundamental principles of programming and basic constructs such as variables, loops, conditional statements, array manipulation, plotting, classes and objects. Each year the class size was usually between 70 and 90 students (the exact number of students can be found in Table \ref{table:total_number_of_students}).

The students embarking on undergraduate study at Imperial College London have a broad range of backgrounds, but these are mostly STEM-based\footnote{STEM stands for Science, Technology, Engineering and Mathematics.} as a result of Imperial College London's focus on STEM subjects. However, no prior knowledge of higher mathematics, programming or computing was assumed. Each year comprised only new students; there were no course `re-takers/re-sitters' from previous years. Of course, the individual student backgrounds and experiences vary from year to year. In general only a few ($\sim$1--6) students had some prior programming experience, sometimes obtained by doing an A-Level\footnote{An A-Level, more formally known as General Certificate of Education (GCE) Advanced Level, is a qualification offered by Further Education institutions in the United Kingdom.} computing course at a Further Education institution prior to embarking on undergraduate study; in the 2010 class there were two students with an A-Level in Computing, in 2012 there was one, and in 2013 there were two, with no students having an A-Level in Computing in other years (see Table \ref{table:alevels} for a full list of student A-Level qualifications by subject). This is typically very different to classes majoring in computer science where a much larger proportion of the intake have some past programming experience, or at least a strong background in logical/algorithmic thinking which will aid them considerably when learning to program. With respect to demographics, UK `home' students (i.e. students who are ordinarily resident in the UK) made up the majority of each intake, and all years featured a larger number of male students, as detailed in Table \ref{table:demographics}.

Whilst Imperial College London is a research intensive university with a worldwide reputation for research excellence, it has made a clear commitment to delivering a world class education for its students; in light of this, teachers across the institution are encouraged to adopt an iterative approach to course design and are offered the freedom to innovate with a view to making teaching as effective as possible. As a starting point for developing the course we adopted the text book
{A Primer on Scientific Programming with Python} \citep{Langtangen_2011}
(and later editions in subsequent years) which is targeted at science
and engineering students with no previous programming experience. The
author of the text book, Hans Petter Langtangen, also provided us with
his collection of slides based on the book which he had developed for his
own introduction to programming course at the University of
Oslo. While the core topics and structure remained largely the same
throughout the years that the course was run, both the teaching medium
and methodology was changed radically in response to student
performance and experience.

Put briefly, our initial approach was to teach in a traditional way: a
3 hour lecture block to cover material, with students being given
exercises at the end of the lecture as coursework. It quickly became
apparent that this approach would require a full additional 3 hour block
timetabled in the computer lab with tutors to support them doing the
exercises. Not only did this put additional pressure on the timetable, it also resulted in the poorest student performance
and experience of the five years of teaching the course.

In subsequent years many changes were made to the course in response
to the student experience which were guided by the experience of the teaching
community and pedagogy literature. Our current approach delivers
satisfactory learning outcomes, and focuses on using the IPython Notebook
software \citep{PER-GRA:2007} with a blended learning approach; this is largely inspired by the
teaching practices promoted by the Software Carpentry organisation (www.software-carpentry.org) which teaches basic computing skills to scientists \citep{Wilson_2006, Wilson_2014}. Currently the 3
hour block is broken down into a series of 10--15 minutes of lecturing
to establish the context and motivation of the current topic. These
include live examples which are worked out and discussed. Between
these mini-lectures are practical exercises where the students are
allocated approximately 30 minutes to work on a few exercises related to the mini-lectures, with
teaching assistants and peers providing support. The IPython Notebook
software neatly integrates core course content written in the Markdown
language \citep{Gruber_2004} with Python code that can be run interactively within the
same document.

In this paper, we report on our experience and the impacts of changing our teaching methodology on the performance of the undergraduate geoscience students. Of course, it can be rather difficult to directly attribute improvements in student learning outcomes to any one specific change in teaching methodology. Each year's student cohort is different and there could be other factors involved, such as variations in the backgrounds of the teaching assistants each year. However, in our study we rely on the fact that we kept the exam format and difficulty level consistent throughout the years. Furthermore, our experiences suggest that if we do not apply any particular methodology then some students perform poorly whilst some do very well, because of the broad range of backgrounds. On the other hand, if we do apply particular pedagogical techniques, a consistent positive learning outcome is achieved.

The ``\nameref{sect:data}'' section gives details of the data collected for evaluating student learning and effectiveness of the course. The ``\nameref{sect:methodology}'' section presents the various design aspects of the current (2014) state of the course, and the teaching practices employed. For each aspect of the course we also contrast with previous years by detailing what we changed each year, why we did this, and the justification for why we believe such changes have helped form an efficient teaching and learning methodology\footnote{Here we define an `efficient methodology' as an approach to teaching that (1) not only delivers the course material within the time and resource constraints, but also (2) improves the students' programming knowledge and skillset, and (3) effectively achieves all the desired learning outcomes and course objectives.}. Finally, some closing remarks and recommendations for adopting the current course methodology elsewhere are given in the \nameref{sect:conclusion}.

\section*{Data Collection}\label{sect:data}
This section describes the types of data that we have collected with respect to student performance and feedback, and the techniques used to obtain this data.

\subsection*{SOLE: Student OnLine Evaluation}
SOLE\footnote{http://www3.imperial.ac.uk/registry/proceduresandregulations/surveys/sole} is the central Student OnLine Evaluation tool developed and maintained by Imperial College London and the Students' Union. The tool enables
students to give their view on their lecturers and modules each year, and also facilitates the assessment of the quality of their degree programme. All surveys are
anonymous and are run at the end of each term. It surveys all
undergraduate students on their modules and the lecturers who have taught those
modules during the term that is finishing. The SOLE module/lecturer
evaluation consists of two sets of questions asked per module and per lecturer.

As part of the annual review of the department for Quality Assurance and
Enhancement purposes, a report is written summarising key issues from SOLE.
Individual staff give feedback to the students, often in the same academic
year, and at the start of the next time the course is taught --- in particular
highlighting issues raised by the students and adjustments made to resolve the
issue. For example, ``students commented that not enough examples were discussed
in-depth so now we have significantly increased the number of examples walked
through in lectures''. A summary of the main outcomes is then sent to each
individual student to complete the feedback loop.

It should be noted that this feedback only represents the students' perceptions of the course and the lecturer, and is not a measure of actual learning.

\subsubsection*{Response rate}
In the Department of Earth Science and Engineering, since 2009, the response
rate for SOLE is between 99 to 100\% thus reducing the risk of sample bias. This gives a complete picture when all
students take the time to complete the survey and give their views. The survey
for the `Introduction to Programming for Geoscientists' course is completed during week 7 out
of 8 classes. There is a strong culture of communication and partnership within
the department. Students welcome the opportunity to provide feedback on their
modules. They see the outcomes of each SOLE survey and know that their
comments make a difference to the future. Students are given an induction in
their first year about how SOLE works and reminded of this annually.  

\subsubsection*{Changes to survey questions/scoring criteria}
An inconsistency in the survey criteria occurred in 2012; the number of criteria was reduced in the hope it would boost response rates across the university as a whole. The survey was reinvented for subsequent years when it was clear that the reduction in criteria did not impact response rates. In addition, two new criteria (``I have received helpful feedback on my work'' and ``The content of the module is intellectually stimulating'') were added, while two criteria (``The organisation of the module'' and ``The support materials available for this module'') were removed.

As with the above, reflecting up-to-date pedagogic thinking, other minor changes of wording appeared. For example in 2010--2012 students were asked to rate the structure of the \textit{lectures} or \textit{teaching sessions}, whereas in 2013 and 2014 they were asked to rate the structure of the \textit{module}. This word replacement reflected the wider nature of the survey from traditional `talk and chalk' lectures.

One other change was the wording of the response, again to reflect up-to-date learning terminology in the UK: ``Very Good (A), Good (B), Satisfactory (C), Poor (D), Very Poor (E), No Response (F)'' were replaced by ``Definitely Agree (A), Mostly Agree (B), Neither Agree or Disagree (C), Mostly Disagree (D), Definitely Disagree (E), Not applicable (F)''. The scoring system used to interpret student feedback in a quantitative fashion in the \nameref{sect:methodology} section is given by

\begin{equation}
   S = \frac{\sum_{i=A}^{E} w_in_i}{\sum_{i=A}^{E} n_i}
\end{equation}

where $S$ is the score for a given criterion, $w_i$ is the numerical value assigned to response $i$ (responses A to E are given scores that scale linearly such that $w_A$ = 2, $w_B$ = 1, $w_C$ = 0, $w_D$ = -1 and $w_E$ = -2), and $n_i$ is the total response count for response $i$.

The same scoring system is used throughout the 5 years the course was run, despite changes in the wording of the responses. An average score of -2 to -1.5 indicates ``Very Poor'', -1.5 to -0.5 indicates ``Poor'', -0.5 to 0.5 indicates ``Satisfactory'', 0.5 to 1.5 indicates ``Good'' and 1.5 to 2 indicates ``Very Good''. 

\subsection*{Examination results}
The programming course had no marked term-time coursework component, and was instead graded based on a two-hour end-of-term examination. In 2010 this involved answering a series of programming-related questions by hand-written word (in the computer lab), while in the later years it involved completing a series of practical exercises and submitting the Python source code files at the end of the exam. Anonymised mark distributions are presented along with the SOLE feedback in the \nameref{sect:methodology} section.

\section*{Current Curriculum and Instruction Methodology}\label{sect:methodology}
In this section we will describe the current state of the course's instruction methodology, and explain the advantages of our approach. We then compare and contrast this with the previous years that the course was run, describing the changes that were made, and give evidence for why these changes were successful. This evidence comprises trends seen in the SOLE data, student comments, and the exam marks. We also explain what changed between the years to cause an improvement in the marks and comments, and how it affected the learning outcomes.

\subsection*{Blended learning approach}
The current state of the course adopts a blended learning approach, following many of the pedagogical methods used by the Software Carpentry organisation \citep{Wilson_2006, Wilson_2014}. Blended learning essentially brings together different modes and techniques of content delivery \citep{BonkGraham_2006, Friesen_2012}. Concepts and ideas are introduced via face-to-face interaction to build a solid theoretical knowledge base. A computer-mediated environment is then used to help apply this knowledge in a practical sense, enhance the students' learning, and develop the necessary skills. Our experience has found that this is an example of a highly efficient methodology since, much like learning to swim or learning a new spoken language, the students' time is most effectively applied to developing their programming skills through practice whilst still having the face-to-face component to acquire the necessary theoretical/background knowledge. Indeed, it has already seen success in other computer programming classes (see e.g. \cite{Boyle_etal_2003}).

In the case of our programming class, the students are expected to read the lecture notes beforehand. The 3-hour-long workshops are divided up into intervals; since empirical evidence has shown that the average adult student can only maintain focus for approximately 15-20 minutes \citep{MiddendorfKalish_1996}, each interval comprises approximately 10 minutes of lecturing to establish the context of a particular section of the lecture notes, followed by a period of time for the students to complete the exercises individually with the lecturer and teaching assistants on-hand to offer help. Empirical evidence has also shown that students who see worked examples before attempting exercises by themselves are better able to tackle future problems \citep{Guzdial_2015}. With this in mind, worked examples are presented during the short lecture before the actual workshop session begins where students can attempt new problems on their own with support on hand if they need it.

At the pre-university level, many students are used to a learning environment in which traditional passive learning and `note-taking' classes are the norm. The student spends the day in a passive state, recording the knowledge being delivered to them by the teacher, often by taking notes directly from the teacher's blackboard. The student is expected to absorb this knowledge and regurgitate it for examination purposes, but they have little responsibility over their own learning. We found that this expectation of a passive, teacher-led environment can therefore be very difficult to change as the student starts university for the first time, particularly as this is still the dominant method of lecturing in most universities. This is a particular issue here as this course is taught in the first term of their first year. Reassurance, explicit justification and a brief explanation of the blended learning approach are therefore given in the very first lecture of the course.

The blended learning methodology can potentially be applied to many other fields of study, not just to programming classes. Indeed, blended learning has seen success in the teaching of biology \citep{YapiciAkbayin_2012}, finance/accounting \citep{Dowling_etal_2003}, and human anatomy \citep{Pereira_etal_2007}, for example. However, while the face-to-face delivery will be similar (i.e. lecturing with slides or a blackboard), the exact form of the technology-based media may be different in the case of non-programming classes. In the case of an electronics class, for instance, the practical exercises may involve designing a circuit rather than writing a program.

\subsubsection*{Comparison with previous years}
In contrast to the initial years that the course was run (2010--2013, in which students were taught through two-hour-long lectures with the use of slides), blended learning was extremely successful and significantly boosted student performance. 

The initial passive lecturing approach was more in line with student expectations. This is illustrated by the 2010 SOLE feedback, with 94\% of students giving a rating of ``Satisfactory'' or higher for the category ``The structure and delivery of the lectures''. Furthermore, the lecturer/lecturing style was not the main trending issue in that year's student survey (see Figure \ref{fig:data_2010}c), and the ``Structure/delivery'' of course material scored a relatively high (`Good') mark of 1.11 based on the SOLE feedback (see Table \ref{table:scores_2010}), as did the course in general (as shown by the `Good' mean combined score, i.e. the mean of the module and lecturer scores, of 1.24 in Figure \ref{fig:data_2010}b). Similarly, a high mean combined score was achieved in 2011 (see Figure \ref{fig:data_2011}b), with a `Good' score of 0.65 for ``Structure/delivery'' (see Table \ref{table:scores_2011}). This standard approach of lecturing for 3 hours may be seen as an efficient methodology from a material dissemination point-of-view \citep{BeardHartley_1984} (i.e. delivering a large amount of material to a high number of students all at once), but is not necessarily effective at achieving learning outcomes \citep{Ramsden_1992, Isaacs_1994}. Indeed, this proved to be the case, with the 2010 mean exam mark of 50.5\% reflecting this (see Figure \ref{fig:data_2010}). In 2011 it was felt that the addition of an extra 3 hour practical session (whilst still covering the same amount of course material) helped to push mean grades to 68.9\%. However, this was not sustainable for future years. In addition, students who could not keep up with the pace of the lectures during 2010 and 2011 quickly lost track (e.g. one student commented ``the lecture powerpoints are sometimes delivered too quickly and then cannot understand the lecture from then on.'' (2011)); this was the biggest trending issue based on the student feedback.

In 2012, a set of lectures were recorded and uploaded to YouTube (www.youtube.com). The YouTube videos were approximately 10--15 minutes long, in an approach largely inspired by online teaching resources such as the Khan Academy (www.khanacademy.org) and Massive Open Online Courses (MOOCs) \citep{YuanPowell_2013}. The students could watch these videos in class at their own pace and complete the exercises within the three-hour timeslot. However, we found that the students then felt unsupported by this approach as they were simply not used to such a flipped classroom approach in which they have much greater responsibility over their own learning: ``I think that it would help if the course was actually taught - at the moment the way its structured there is no need for a teacher to be in the room as he does not cover any content within the lessons.'' (2012), ``I found the teaching to be very impersonal but maybe that is because programming is something you have to learn for yourself.'' (2012), ``He doesn't lecture'' (2012), ``People end up being able to do things but not having a clue why as nothing is explained.'' (2012). This was also reflected in the SOLE feedback and course scores; a much higher proportion (28\%) of first year students feeling that the ``The structure and delivery of the lectures'' was less than satisfactory, a relatively low score of 0.38 for ``Structure/delivery'' (see Table \ref{table:scores_2012}), and a fairly broad distribution of marks with the mean being 60.3\%. One of the reasons for the lack of engagement in this approach may in part be related to the place in the curriculum. As the course is in the first term of their first year in University, not all students had developed their independent learning skills. Furthermore, evidence provided by effect sizes \citep{Hattie_2008} has shown that web-based learning only has a small positive influence on learning relative to the traditional classroom environment. Despite this negative feedback from the students the learning outcomes were much better than the first year the course was run.

Because of the lack of constructive impact, we decided to no longer use online videos and instead opted for traditional lecture notes in 2013. A short lecture of 20--30 minutes was delivered at the beginning of the class, but students were expected to have read the notes before the class began in order to maximise the amount of time that could be spent on the practical exercises that followed. However, while a few students understood the need for this self-preparation (``He gives you the responsibility to succeed and the tools to do so, I do not see that he can do any more unless he were to baby feed us which isn't why we are at university.'' (2013)), the majority of the student feedback regarding expectations from the lecturer still amounted to `the lecturer is not lecturing us': ``Being shown a lecture and then expected to complete an exercise with little teaching is difficult.'' (2013), ``The lecturer should teach us the content of the course, instead of us having to read it off notebook with limited explanation to go with those examples.'' (2013), ``More lecture based learning. Not enough explanation of lecture notes and poor quality lecture notes'' (2013), despite reassuring them that it was more effective for them to read the lecture notes themselves at their own pace and spend the majority of the time doing the exercises, rather than the lecturer spending most of the available time reading the lecture notes out to the students. This was also reflected in the lecturer scores, with a negative score of -0.06 for the ``Explained material'' category (see Table \ref{table:scores_2013}) contributing to a relatively low mean combined score of 0.58 for the whole course, as illustrated in Figure \ref{fig:data_2013}b. Despite this, our data suggested the change to a flipped classroom environment was beneficial to the students' performance; this is reflected in the mark distribution which is skewed towards the higher end of the spectrum, and the mean course mark of 74.5\% was considerably higher than previous years, as shown in Figure \ref{fig:data_2013}.

Finally, in 2014 when the blended learning approach was adopted (and, crucially, justified to the students), it was clear that students understood the need for such an approach when learning to program. This was clear from the SOLE feedback: ``I agree with [the lecturer] that we gain much more from practicals than from being lectured'' (2014), ``[I] understand the need for self teaching'' (2014). The score for course structure and delivery increased to an all time high of 1.26, yet the score for explanation of the material also remained high at 1.11 (see Table \ref{table:scores_2014}) through breaking the 3-hour workshop down into individual small lectures and reassuring the students that the teaching approach taken was beneficial, albeit unlike what they were used to. The mean combined score of 1.4 for the whole course was also relatively high compared to other courses run by the department that year (see Figure \ref{fig:data_2014}b). At the same time, the mean exam mark did not change considerably. The overall mark distribution looked much like the one of 2011, but required just 3 hours of teaching time per week compared to the (unsustainable) 6 hours allocated in 2011.

\subsection*{Tailoring of material, exercises, and pace}
Students majoring in computer science will immediately see a need for, and typically have a strong interest in, learning to write computer programs. In contrast, our experience with teaching geoscience students has shown that it is crucial to motivate the need for programming from the very first lecture, since many often feel that they are being forced to do something that was not relevant or worthwhile in their undergraduate syllabus. To that end, the course includes several exercises that are tailored towards geophysical scenarios. For example, students are asked to create a program which reads in seismic data from a file and locates the earthquake of the largest magnitude. Another exercise involves reading a tidal gauge data file supplied by the British Oceanographic Data Centre, plotting the tide level against time, and then using that to spot the tidal constituents. In addition, we use our own computational-based research to further justify why students would want to develop programming skills by showing them simulations of volcanic eruptions and seismic wave propagation produced by computer programs.

It is also important to make the learning experience as interactive as possible. We therefore encourage the students to discuss problems with one another to aid peer-learning. One exercise is particularly successful at engaging the students; the exercise tasks them with creating a `Battleship' game in Python, which they later play against their neighbour. Not only is this an enjoyable exercise, but it is also challenging enough to bring together many of the topics the students learn in the course. Furthermore, we observed that this can promote several instances of `rubber duck debugging' \citep{HuntThomas_1999}; students manage to spot inconsistencies between what their program is actually doing and what they expect it to do, simply by walking through their program with one of their peers without that peer necessarily saying anything at all. This form of peer assessment proved to be an efficient way of testing the application of knowledge while allowing a degree of peer instruction as students could correct each other if necessary. The informal nature of the game-based scenario promoted independence, control, active engagement and enjoyment which have been held up as key principles in effective teaching and learning in higher education \citep{Ramsden_1992}.

Finally, in order to facilitate effective learning we needed to manage the pace and cognitive load being placed on students without diluting the quality or academic rigour of the course. The course material was therefore refined each year following reviews of the course whilst still keeping the core learning outcomes in place.

\subsubsection*{Comparison with previous years}
Throughout the years that the course was run, there was always initial resentment from a minority of students who, as geoscientists, felt that they were being forced to do something that was not useful (e.g. ``I don't understand how it is related to the other stuff we study in geology.'' (2012)). It was therefore crucial to underline the importance of programming skills from the first lecture with the hope this would have a knock on effect on the students' motivation to learn. This was partially accomplished by citing real geophysical applications that involve software development. This was reflected in the lecturer-specific feedback given towards the end of the term; in 2010 and 2011 respectively, 97\% and 92\% of students thought that ``The interest and enthusiasm generated by the lecturer'' was at least satisfactory. In 2013 and 2014 respectively (when the SOLE rating options changed), 89\% and 99\% of students either had no opinion, mostly agreed, or strongly agreed that ``The lecturer generated interest and enthusiasm''. The student comments also show how the resentment was overcome once the student's own resistance to programming was mitigated through motivation: ``I feel that although at first I despised programming this was due to my own block against the subject.'' (2013).

When the course commenced for the first time in 2010, the exercises in the book by \cite{Langtangen_2011} were considered. These were largely subject-independent and some students therefore found it hard to relate these exercises to problems that they would deal with as geoscientists in the real world; for example, in 2013 one student ``felt that it took some time to get the general picture, the lecturer would better help the students if he applied the idea of programming to real life''. The tailoring of the exercises towards geophysical scenarios was therefore a change that was well-received by the students.

Many of these exercises required the implementation of equations and mathematical functions (e.g. implementing the Heaviside function). The students were tasked with translating these formulae into code. Although a very high proportion of the students had strong backgrounds in mathematics at A-Level (see Table \ref{table:alevels}), it is not a prerequisite and the students were not required to understand the mathematics behind the formulae. Nevertheless, the presence of mathematics concerned some students and demotivated them; anonymous student comments that highlight this include ``Some of the exercises assumed A level Maths as a prerequisite which was not appropriate for all students and very distressing for some.'' (2013), ``you have to figure out the Maths before you can even start to programme'' (2013), ``a lot of the exercises assume a previous understanding of some complicated maths which adds to the confusion of the programming itself.'' (2013). It was therefore necessary to provide reassurance and demonstrate that an in-depth knowledge of the derivation and use of the mathematical formulae is not required for them to be able to complete the exercise; for example, showing that a finite sum of sine functions can be implemented using a for-loop.

Particularly during the first few years that the course was run, students largely felt overwhelmed with the amount of material covered (``At the start there was too much to do.'' (2012)) and that many lectures had to be rushed at the end due to lack of time (``Far too much content to fit into 8 lectures!!!!!! Not once did we finish a lecture.'' (2010), ``The amount of material is covered in a - sometimes - too short amount of time.'' (2011)). This is clearly visible in Figures \ref{fig:data_2010} and \ref{fig:data_2011} where the largest (negative) trending topic in 2010 and 2011 concerned the pace of the lectures. The students responded positively to the refinement of the course in 2012, with one student commenting that ``The content of the programming for geoscientists was perfect when cut down to only 6 parts.'' (2012).

\subsection*{Practice}
We found that the short live lecture in a flipped classroom environment followed by the `bite-size' chunks of practical exercises, was a highly effective and efficient way of developing programming skills. The context, background material and example code are all covered just before the students attempted the associated exercises. The material is therefore fresh in their minds and gives students a firmer foundation on which to practice their skills in the particular topic under consideration. Furthermore, any questions that students have can be asked at the classwide-level during the live lecture, thereby addressing queries that may be shared by many students in one go.

\subsubsection*{Comparison with previous years}
In comparison with the 2010 run of the course, the students initially spent two-hours being lectured new material, and then given one hour to complete a range of practical exercises. However, students were struggling to complete these exercises in the time allocated. The course greatly improved a few weeks into the term after extending the practical sessions by 3 hours (on the same day), as reflected in the students' feedback: ``This course definitely improved after the reading week\footnote{During `Reading Week', students have no lectures and are expected to concentrate solely on coursework.} when they introduced workshops in the afternoon with the demonstrators.'' (2010), ``I believe the new structure of having 3 hours of GTA help and worked through solutions help.'' (2010), ``The introduction of a separate practical session was good.'' (2010). However, at least one student believed that this was ``too little too late'' (2010). Furthermore, it was likely that students became exhausted since the majority of that day involved a constant focus on programming.

When the course commenced in 2011, a stand-alone 3-hour practical workshop was held each week so that the students could focus entirely on the exercises; the students responded positively to this: ``The workshops are absolutely essential. Well run and incredibly helpful.'' (2011) and it appears that this resulted in a more positive skewness in the mark distribution for 2011 as shown in Figure \ref{fig:data_2011}. However, the extra 3 hours of allocated practical time was unsustainable for future years, and a more efficient means of content delivery and practice was therefore required.

The use of YouTube videos in 2012 permitted the students to review the course material before or during the practical session, thereby leaving more time for practice in the computer lab with the support of the Graduate Teaching Assistants (GTAs) and lecturer. However, if a student did not understand a concept in the videos then more support time was taken up explaining concepts at an individual level, instead of practising and obtaining help with the exercises. In comparison, the change to a flipped classroom environment followed by the longer practical session in 2013, and the `bite-size' chunks of practical exercises in 2014, was a far more effective method of programming skill development as shown by the improved examination marks and learning outcomes in Figure \ref{fig:data_2014}.

\subsection*{Technical Considerations}
All lecture and examination material is currently written in the Interactive Python (IPython) Notebook format which has proven to be an extremely effective learning environment as it dispensed with the cognitive load of learning an editor, Integrated Development Environment (IDE) or the command-line interface. It allows students to write and execute their programs in amongst the lecture notes themselves so that everything `flows' and they have all the course material in one place. It also facilitates the running of live examples with the lecturer since the students can more easily follow along.

In order to open the IPython Notebooks, we use Python distributions which run locally on the Microsoft Windows-based lab computers. Students can choose between Anaconda \citep{Anaconda} and Enthought Canopy \citep{Enthought}. The students are comfortable with this simple browser-based environment, and it also removes a considerable amount of complexity and setup time. In addition, the ease of installation and the cross-platform nature of the environment (running on Microsoft Windows, Mac OS and Linux) means that it is simple for the students to install these distributions on their own computers for use outside of the computer lab.

Note that, since the students' Python code is combined within these `notebooks', this approach comes with the caveat that it is not possible to easily apply updates or corrections to the course material after the lecture. However, once the course material was revised throughout the years and became stable/mature enough, this was no longer a significant issue.

\subsubsection*{Comparison with previous years}
Different learning environments were considered throughout the 5 years that the programming course was run. Integrated Development Environments (IDEs) were discounted from the outset because of their potential for causing high cognitive load on novices in introductory programming courses \citep{Pears2007}. For the first four years, students accessed a central server running the Ubuntu Linux operating system in order to write and execute their programs. During the first week of term before any programming was taught, the students were given a primer on basic Linux command-line tools by the local system administrator. While knowledge of Linux commands was not being examined, students had to focus not only on learning to program but also deal with the challenge of getting used to a new learning environment. The additional level of complexity, which could be interpreted as placing extraneous load on the learners, sometimes led to frustration and demotivation, particularly when some technical difficulties could not be readily resolved by the teaching assistants: ``Not enough work on linux'' (2010), ``Started quite badly. Most students had never touched linux or python.'' (2010). Most students were accustomed to graphical interfaces in their day-to-day computer use as opposed to command line interfaces. The other complication was that they could not (or found it difficult to) install Linux or Python on their own computers, thereby preventing self-study outside of the computer lab.

In 2010, students wrote their programs in a console-based text editor such as Nano \citep{nano_2009} and executed the program with the Python interpreter directly at the command line. Students had to continually switch between the editor and the command line, and it was clear that writing and running stand-alone program files in this way required a considerable amount of additional expertise which placed an extraneous cognitive load on the students. To help remedy this in 2011 and 2012, students adopted the Interactive Python (IPython) tool which simplified the process of seeing a program's results and lowered the turn-around time for debugging considerably. Additionally, in 2013, the IPython Notebook format was adopted (see above). The Git version control system was used in an attempt to apply any updates/corrections to the lecture notes as gracefully as possible, and also gave the students an insight into using version control to manage their work. However, this turned out to be somewhat counter-productive as merge conflicts frequently had to be resolved manually by the teaching assistants which lowered the confidence the students had in the system they were using, and added to the number of commands the students had to remember to download the latest revision of the lecture material. The majority of students in the class struggled to cope with both Linux and Git which were completely new to them.

\subsection*{Formative feedback / the `sticky note' system}
The `sticky-pad' system, promoted by the Software Carpentry organisation, was used in every workshop since 2014. This almost counterculture low-tech system proved both to be incredibly powerful and popular with students for obtaining support and providing feedback that was acted on promptly. Essentially, each student is given a red and a green sticky note at the start of every lecture, and they are used in two ways.

In the first case, the sticky notes act as a status indicator when completing practical exercises; the student sticks up the red note to indicate that they need assistance, which eliminates the need to wait passively with their hand raised or similarly trying to get the attention of the lecturer or GTAs (the need to hold up their hand whilst waiting for a GTA to become available had been a frequent point of frustration). This in turn increases class productivity. When the student has finished a set of practical exercises, they stick up the green note to let the lecturer know the overall progress of the class.

In the second use case, the sticky notes act as `exit tickets'; students must leave one piece of positive and negative feedback on the green and red notes, respectively, before exiting the computer lab at the end of the class. This helped to identify immediately if there were issues arising in the course so they could be resolved before the next lecture, and how the class was finding particular aspects of the course. When students struggle on exercises they tend to complain a great deal via the anonymous sticky notes, so changes in class progress in response to feedback were more easily identifiable with this technique.

\subsubsection*{Comparison with previous years}
A great deal of positive SOLE feedback resulted from the use of these sticky notes, as shown by the large positive proportion (51.4\%) of `support'-related comments in Figure \ref{fig:data_2014} and the individual SOLE comments such as ``Sticky notes work well'', ``WE LOVE RED AND GREEN POST IT NOTES!!!!'' and ``Love the post it notes.''. Its effectiveness is also demonstrated through the increase from a score of 0.47 in 2013 to 1.32 in 2014 for the ``Feedback'' criterion (see Tables \ref{table:scores_2013} and \ref{table:scores_2014}).

\subsection*{Teaching and learning support}
It is an important, albeit often challenging task, to bring all the students up to a similar standard within a fixed period of time. The learning pace of students varies significantly for many reasons. This is particularly true for a first year course where the students come from a diverse range of educational backgrounds. Graduate Teaching Assistants (GTAs) play a critical intervention role here\footnote{All teaching assistants in the department receive training in pedagogical techniques and marking before they are permitted to help out in undergraduate courses.}. They help to clarify concepts, support progress through exercises (identifying key difficulties), provide instant feedback on the students' work, and resolve practical computing issues. We therefore ensure that a high number of GTAs are present in each workshop. The total number of GTAs who assisted each year is given in Table \ref{table:gtas}. Typically there are between 8 and 10 GTAs per workshop in the current version of the course, yielding a student to GTA ratio of at least $\sim$10:1.

\subsubsection*{Comparison with previous years}
While the class could be large (typically 70--90 students each year in this case) it proved critical to have a low ratio of students to GTAs so that the students did not experience long and unproductive/demoralising waiting times for one-to-one support. This issue consistently featured strongly in student feedback; highly negative when there was a shortage of teaching assistants (typically 4 or fewer), high degree of satisfaction and praise for the teaching assistants when there were enough (typically 8 -- 10 from 2013 onwards). This was an issue in 2012 when, exceptionally, the course was run for both first and second year students concurrently. This was necessary to transition the course from being a `second year only' course to being a `first year only' course in later years. Two computer labs were used at the same time, but the number of available teaching assistants had to be spread out; this resulted in comments such as ``Could do with more teaching assistants'' and ``Need more demonstrators''. However, in the other years (especially in 2013 and 2014) when just first year students were being taught, the ratio of students to GTAs was low and the teaching assistants received a great deal of praise, such as ``Good demonstrators and feedback as well. Don't have to wait very long if there's a problem, and usually solved quickly.'' (2010), ``GTAs are particularly helpful.'' (2013), and ``Very helpful GTAs.'' (2014).

\section*{Conclusion}\label{sect:conclusion}
The successful development of an introductory computer programming course represents a significant challenge, particularly when targeting undergraduate students with little or no computing background and outside a mainstream computer science degree programme. This study has outlined the influence of different methodologies on student perceptions of learning, and what appears to be a positive impact on student examination results, over the course of 5 years. Our findings reinforce the evidence in the literature that flipped classrooms and blended learning approaches are much more effective at teaching programming than traditional passive lecturing. However, we accept that our study would require further, more rigorous investigation, potentially using a quasi-experimental design, to demonstrate learning gain.

The traditional passive lecturing style that featured in the early years of the course (2010 and 2011) was more in line with what students were used to, yet was ineffective at developing the students' skillset. The video lectures that were implemented in 2012 allowed students to work through material and review it at their own pace, but a high student--GTA ratio and the lack of traditional lecturing style was a concern for many students. In 2013 when a flipped classroom approach was used, students did not feel like they were being taught, leading to a lack of confidence both in themselves and in the course. It is problematic to make a direct link between these interventions and end-of-term examination performance, but we believe this demonstrates the rewards of a `learning through doing' approach which would benefit from further exploration via future studies. It is our belief that we have now converged on a successful teaching strategy through the use of blended learning and formative feedback, which featured in the 2014 run.

That said, many challenges surrounding the teaching of programming still remain. As class sizes grow, some degree of automated marking would be beneficial for the lecturer and teaching assistants. However, this is technically difficult. While it is possible to automatically run all the students' programs and determine whether they produce the correct answer, such an answer must usually be something simple, like a single integer, rather than a plot, for example. Furthermore, much like systems such as flake8 \citep{Flake8_2015} which check Python coding style compliance with the PEP 8 standard \citep{PEP8_2001}, it is possible to automate checks such as `Has the student added docstrings for each function?'; however, determining whether these docstrings or comments are actually useful or not presents a much bigger challenge.

There is a large amount of pedagogical research and knowledge regarding how people learn and how best to teach programming skills. Despite this, many courses still follow the more traditional passive lecturing style, since change may be viewed as a risky process from an institutional point-of-view. It is therefore hoped that the findings presented in this paper will inform and encourage departments and educators to reconsider their existing approach to teaching programming.

\subsection*{Recommendations for Adoption}
The various aspects of our course's instructional design can be readily applied to other courses. For example, the sticky note system is not just applicable to computer programming, and can indeed be applied to most courses that involve the completion of in-class exercises and require one-to-one help from the lecturer/GTAs. However, the appropriateness of the technical aspect will need to be considered carefully; we found that the IPython Notebook is an excellent environment to teach programming, but for practical exercises not involving computer code, a different practical setup may have to be designed. For example, for mechanical engineering design problems, a Computer Aided Design (CAD) package could be a more appropriate learning environment. For pure mathematics an interactive symbolic algebra package such as Maple may be desirable (although symbolic algebra can also be handled by readily-importable Python modules such as SymPy \citep{Joyner_etal_2012}).

The course material is freely available under the Creative Commons Attribution 3.0 Unported (CC-BY 3.0) licence and can be tailored to an individual setting. It can be downloaded from \texttt{https://github.com/ggorman/Introduction-to-programming-for-geoscientists}.

\subsection*{Outlook}
Although the findings in this paper provide a valuable insight into the methodologies used to teach computer programming, we acknowledge that the limitations in the type and quantity of the available data may be detrimental to the reproducibility of our findings. We therefore believe that it would be beneficial to conduct a more rigorous, quasi-experimental study in the future, with a more formal data collection plan from the outset.

Given that using a blended learning methodology throughout the last two years that the course was run featured consistently high performance, we hope that similar performance would be maintained in future classes. However, a much more rigorous study would be required to confirm this. With such a study we could also potentially investigate whether there is any correlation in, for example, the number of students with a Mathematics or Computing A-Level and examination performance. Although we don't envisage any significant effects of a change of lecturer on student performance (as long as the methodology was applied consistently and the new lecturer had the appropriate background and experience), this would be another avenue of investigation to consider, once again requiring appropriate evidence to back up this statement. On the other hand, we found that some GTAs were mentioned by name more than others and given high praise in the student feedback; it is unclear whether the unavailability of these particular GTAs during some weeks (due to the rota system) affected performance since student satisfaction is not a measure of actual learning. However, this is something that will be looked at more closely in later work.

For the purposes of demonstrating how student examination responses were graded, and to illustrate that the responses were of a consistent quality throughout the years, a revised study would also record anonymised examples during the data collection phase. We plan to consider both high and low quality responses for comparison.

Further improvements in the software supporting the students' learning (in this case, the IPython Notebook) may potentially affect student performance. For example, the version used in 2014 was prone to crashing when infinite `while' loops occurred. Students were able to seek assistance swiftly from the GTAs when such technical difficulties occurred but sometimes caused frustration, especially if the IPython Notebook had to be restarted and part of the student's work was lost. If such software issues were to be resolved, this may boost the students' satisfaction and confidence in the course. However, once again, a more rigorous study would be needed to show whether this positively affects student performance, since improved student satisfaction does not necessarily yield learning gain.

Other courses that form part of the geoscience curriculum would also be considered, as it could act as a way of measuring how well information is retained from the introductory programming course. For example, in 2014 another course took place the following term that required the use of Python. More specifically, it focussed on the application of Python's numerical and scientific libraries NumPy and SciPy to perform statistical analysis of geoscientific data. This has since been replaced (for the 2015 class, outside the timeframe of the current study) by a course that concerns the Python implementation of numerical methods to solve systems of equations modelling geophysical phenomena. Both of these courses aim to further the students' education in computational geoscience.

\section*{Acknowledgements}\label{sect:acknowledgements}
We would like to thank all the undergraduate geoscience students that we have taught over the years in the Department of Earth Science and Engineering at Imperial College London for providing constructive feedback on the course as it evolved. We are also extremely grateful for all the hard work and support received from the GTAs over all the years teaching this course. Special thanks go to Hans Petter Langtangen for giving permission to use content from his book and course notes, and to Greg Wilson of the Software Carpentry project for training and useful discussions on evidence-based pedagogy for programming. The feedback from the anonymous reviewers was warmly received and greatly improved this article.

This work was carried out under the approval from the Education Ethics Review Process (EERP) at Imperial College London.

\bibliographystyle{plainnat}
\bibliography{teaching-python-to-geoscientists}

\begin{thebibliography}{47}
\providecommand{\natexlab}[1]{#1}
\providecommand{\url}[1]{\texttt{#1}}
\expandafter\ifx\csname urlstyle\endcsname\relax
  \providecommand{\doi}[1]{doi: #1}\else
  \providecommand{\doi}{doi: \begingroup \urlstyle{rm}\Url}\fi

\bibitem[Aagaard et~al.(2007)Aagaard, Williams, and Knepley]{aagaard2007pylith}
B.~Aagaard, C.~Williams, and M.~Knepley.
\newblock {PyLith: A finite-element code for modeling quasi-static and dynamic
  crustal deformation}.
\newblock \emph{Eos Trans. AGU}, 88:\penalty0 52, 2007.

\bibitem[Beard and Hartley(1984)]{BeardHartley_1984}
R.~M. Beard and J.~Hartley.
\newblock \emph{{Teaching and Learning in Higher Education}}.
\newblock Harper \& Row, London, 4th edition, 1984.

\bibitem[Beyreuther et~al.(2010)Beyreuther, Barsch, Krischer, Megies, Behr, and
  Wassermann]{beyreuther2010obspy}
M.~Beyreuther, R.~Barsch, L.~Krischer, T.~Megies, Y.~Behr, and J.~Wassermann.
\newblock {ObsPy: A Python toolbox for seismology}.
\newblock \emph{Seismological Research Letters}, 81\penalty0 (3):\penalty0
  530--533, 2010.

\bibitem[Biddle and Tempero(1998)]{biddle1998java}
R.~Biddle and E.~Tempero.
\newblock Java pitfalls for beginners.
\newblock \emph{ACM SIGCSE Bulletin}, 30\penalty0 (2):\penalty0 48--52, 1998.

\bibitem[Bonk and Graham(2006)]{BonkGraham_2006}
Curtis~J. Bonk and Charles~R. Graham.
\newblock \emph{{The Handbook of Blended Learning: Global Perspectives, Local
  Designs}}.
\newblock Pfeiffer, San Francisco, 1st edition, 2006.

\bibitem[B{\"o}sz{\"o}rm{\'e}nyi(1998)]{boszormenyi1998java}
L.~B{\"o}sz{\"o}rm{\'e}nyi.
\newblock {Why Java is not my favorite first-course language}.
\newblock \emph{Software-Concepts \& Tools}, 19\penalty0 (3):\penalty0
  141--145, 1998.

\bibitem[Boyle et~al.(2003)Boyle, Bradley, Chalk, Jones, and
  Pickard]{Boyle_etal_2003}
Tom Boyle, Claire Bradley, Peter Chalk, Ray Jones, and Poppy Pickard.
\newblock {Using Blended Learning to Improve Student Success Rates in Learning
  to Program}.
\newblock \emph{Journal of Educational Media}, 28\penalty0 (2-3):\penalty0
  165--178, 2003.
\newblock \doi{10.1080/1358165032000153160}.

\bibitem[Churcher and Tempero(1998)]{churcher1998java}
N.~Churcher and E.~Tempero.
\newblock Java as a first programming language.
\newblock In \emph{Software Engineering: Education and Practice, International
  Conference on}, pages 390--390. IEEE Computer Society, 1998.

\bibitem[Clark et~al.(1998)Clark, MacNish, and Royle]{clark1998java}
D.~Clark, C.~MacNish, and G.~F. Royle.
\newblock {Java as a teaching language --- opportunities, pitfalls and
  solutions}.
\newblock In \emph{Proceedings of the 3rd Australasian conference on Computer
  science education}, pages 173--179. ACM, 1998.

\bibitem[Close et~al.(2000)Close, Kopec, and Aman]{close2000cs1}
R.~Close, D.~Kopec, and J.~Aman.
\newblock {CS1: perspectives on programming languages and the breadth-first
  approach}.
\newblock In \emph{Journal of Computing Sciences in Colleges}, volume~15, pages
  228--234. Consortium for Computing Sciences in Colleges, 2000.

\bibitem[{Continuum Analytics}(2015)]{Anaconda}
{Continuum Analytics}.
\newblock {Anaconda Scientific Python Distribution:
  https://store.continuum.io/cshop/anaconda/ (last accessed: 16 May 2015)},
  2015.

\bibitem[Donaldson(2003)]{Donaldson_2003}
T.~Donaldson.
\newblock {Python as a First Programming Language for Everyone}.
\newblock In \emph{{Western Canadian Conference on Computing Education}}, 2003.

\bibitem[Dowling et~al.(2003)Dowling, Godfrey, and Gyles]{Dowling_etal_2003}
C.~Dowling, J.~M. Godfrey, and N.~Gyles.
\newblock {Do hybrid flexible delivery teaching methods improve accounting
  students' learning outcomes?}
\newblock \emph{Accounting Education}, 12\penalty0 (4):\penalty0 373--391,
  2003.
\newblock \doi{10.1080/0963928032000154512}.

\bibitem[{Enthought Scientific Computing Solutions}(2015)]{Enthought}
{Enthought Scientific Computing Solutions}.
\newblock {Enthought Canopy: Python Distribution and Integrated Analysis
  Environment: https://www.enthought.com/products/canopy/ (last accessed: 16
  May 2015)}, 2015.

\bibitem[{Environmental Systems Research Institute}(2015)]{ESRI_2015}
{Environmental Systems Research Institute}.
\newblock {ArcGIS Project: http://www.arcgis.com (last accessed 16 May 2015)},
  2015.

\bibitem[Fangohr(2004)]{Fangohr_2004}
H.~Fangohr.
\newblock {A Comparison of C, MATLAB, and Python as Teaching Languages in
  Engineering}.
\newblock In M.~Bubak, G.~D. {van Albada}, P.~M.~A. Sloot, and J.~Dongarra,
  editors, \emph{Computational Science - ICCS 2004}, volume 3039 of
  \emph{Lecture Notes in Computer Science}, pages 1210--1217. Springer Berlin
  Heidelberg, 2004.
\newblock ISBN 978-3-540-22129-6.
\newblock \doi{10.1007/978-3-540-25944-2\_157}.

\bibitem[{Flake8 Development Team}(2015)]{Flake8_2015}
{Flake8 Development Team}.
\newblock {Flake8 project documentation: http://flake8.readthedocs.org (last
  accessed: 16 May 2015)}, 2015.

\bibitem[Friesen(2012)]{Friesen_2012}
N.~Friesen.
\newblock {Defining Blended Learning}.
\newblock Technical report, 2012.
\newblock URL
  \url{http://learningspaces.org/papers/Defining\_Blended\_Learning\_NF.pdf}.

\bibitem[{GRASS Development Team}(2015)]{GRASS_GIS_software}
{GRASS Development Team}.
\newblock \emph{{Geographic Resources Analysis Support System (GRASS GIS)
  Software}}.
\newblock Open Source Geospatial Foundation, USA, 2015.
\newblock URL \url{http://grass.osgeo.org}.

\bibitem[Gruber(2004)]{Gruber_2004}
J.~Gruber.
\newblock {The Markdown language:
  http://daringfireball.net/projects/markdown/}, 2004.

\bibitem[Guzdial(2015)]{Guzdial_2015}
M.~Guzdial.
\newblock What's the best way to teach computer science to beginners?
\newblock \emph{Communications of the ACM}, 58\penalty0 (2):\penalty0 12--13,
  2015.
\newblock \doi{10.1145/2714488}.

\bibitem[Hattie(2008)]{Hattie_2008}
J.~Hattie.
\newblock \emph{{Visible Learning: A Synthesis of Over 800 Meta-Analyses
  Relating to Achievement}}.
\newblock Routledge, 2008.

\bibitem[Hunt and Thomas(1999)]{HuntThomas_1999}
A.~Hunt and D.~Thomas.
\newblock \emph{{The Pragmatic Programmer: From Journeyman to Master}}.
\newblock Addison-Wesley, 1999.

\bibitem[Isaacs(1994)]{Isaacs_1994}
G.~Isaacs.
\newblock {Lecturing practices and note-taking purposes}.
\newblock \emph{Studies in Higher Education}, 19\penalty0 (2):\penalty0
  203--216, 1994.
\newblock \doi{10.1080/03075079412331382047}.

\bibitem[Jacobs and Piggott(2015)]{Jacobs2015}
C.~T. Jacobs and M.~D. Piggott.
\newblock {Firedrake-Fluids v0.1: numerical modelling of shallow water flows
  using an automated solution framework}.
\newblock \emph{{Geoscientific Model Development}}, 8\penalty0 (3):\penalty0
  533--547, 2015.
\newblock \doi{10.5194/gmd-8-533-2015}.
\newblock URL \url{http://www.geosci-model-dev.net/8/533/2015}.

\bibitem[Joyner et~al.(2012)Joyner, \v{C}ert\'{\i}k, Meurer, and
  Granger]{Joyner_etal_2012}
David Joyner, Ond\v{r}ej \v{C}ert\'{\i}k, Aaron Meurer, and Brian~E. Granger.
\newblock Open source computer algebra systems: Sympy.
\newblock \emph{ACM Communications in Computer Algebra}, 45\penalty0
  (3/4):\penalty0 225--234, 2012.
\newblock ISSN 1932-2240.
\newblock \doi{10.1145/2110170.2110185}.

\bibitem[Langtangen(2009)]{Langtangen_2011}
H.~P. Langtangen.
\newblock \emph{{A Primer on Scientific Programming with Python}}.
\newblock Springer Berlin Heidelberg, 2009.

\bibitem[Lin(2012)]{Lin_2012}
J.~W.-B. Lin.
\newblock {Why Python Is the Next Wave in Earth Sciences Computing}.
\newblock \emph{{Bulletin of the American Meteorological Society}}, 93\penalty0
  (12):\penalty0 1823--1824, 2012.
\newblock \doi{10.1175/BAMS-D-12-00148.1}.

\bibitem[Middendorf and Kalish(1996)]{MiddendorfKalish_1996}
J.~Middendorf and A.~Kalish.
\newblock {The ``Change-Up'' in Lectures}.
\newblock \emph{The National Teaching \& Learning Forum}, 5\penalty0 (2), 1996.

\bibitem[Mody(1991)]{mody1991c}
R.~P. Mody.
\newblock C in education and software engineering.
\newblock \emph{ACM SIGCSE Bulletin}, 23\penalty0 (3):\penalty0 45--56, 1991.

\bibitem[{Nano Development Team}(2009)]{nano_2009}
{Nano Development Team}.
\newblock {GNU nano project website: http://www.nano-editor.org/ (last
  accessed: 16 May 2015)}, 2009.

\bibitem[Palumbo(1990)]{palumbo1990programming}
D.~B. Palumbo.
\newblock Programming language/problem-solving research: A review of relevant
  issues.
\newblock \emph{Review of Educational Research}, 60\penalty0 (1):\penalty0
  65--89, 1990.

\bibitem[Pears et~al.(2007)Pears, Seidman, Malmi, Mannila, Adams, Bennedsen,
  Devlin, and Paterson]{Pears2007}
A.~Pears, S.~Seidman, L.~Malmi, L.~Mannila, E.~Adams, J.~Bennedsen, M.~Devlin,
  and J.~Paterson.
\newblock A survey of literature on the teaching of introductory programming.
\newblock \emph{SIGCSE Bull.}, 39\penalty0 (4):\penalty0 204--223, December
  2007.
\newblock ISSN 0097-8418.
\newblock \doi{10.1145/1345375.1345441}.
\newblock URL \url{http://doi.acm.org/10.1145/1345375.1345441}.

\bibitem[Pereira et~al.(2007)Pereira, Pleguezuelos, Mer\'{i}, Molina-Ros,
  Molina-Tom\'{a}s, and Masdeu]{Pereira_etal_2007}
J.~Pereira, E.~Pleguezuelos, A.~Mer\'{i}, A.~Molina-Ros, M.~C.
  Molina-Tom\'{a}s, and C.~Masdeu.
\newblock {Effectiveness of using blended learning strategies for teaching and
  learning human anatomy}.
\newblock \emph{Medical Education}, 41\penalty0 (2):\penalty0 189--195, 2007.
\newblock ISSN 1365-2923.
\newblock \doi{10.1111/j.1365-2929.2006.02672.x}.

\bibitem[P\'erez and Granger(2007)]{PER-GRA:2007}
F.~P\'erez and B.~E. Granger.
\newblock {IP}ython: a system for interactive scientific computing.
\newblock \emph{Computing in Science and Engineering}, 9\penalty0 (3):\penalty0
  21--29, May 2007.
\newblock ISSN 1521-9615.
\newblock \doi{10.1109/MCSE.2007.53}.
\newblock URL \url{http://ipython.org}.

\bibitem[{QGIS Development Team}(2009)]{QGIS_software}
{QGIS Development Team}.
\newblock \emph{QGIS Geographic Information System}.
\newblock Open Source Geospatial Foundation, 2009.
\newblock URL \url{http://qgis.osgeo.org}.

\bibitem[Ramsden(1992)]{Ramsden_1992}
P.~Ramsden.
\newblock \emph{{Learning to Teach in Higher Education}}.
\newblock {Psychology Press}, 1992.

\bibitem[Rathgeber et~al.(2015)Rathgeber, Ham, Mitchell, Lange, Luporini,
  McRae, Bercea, Markall, and Kelly]{Rathgeber2015}
F.~Rathgeber, D.~A. Ham, L.~Mitchell, M.~Lange, F.~Luporini, A.~T.~T. McRae,
  G.-T. Bercea, G.~R. Markall, and P.~H.~J. Kelly.
\newblock {Firedrake: automating the finite element method by composing
  abstractions}.
\newblock \emph{Submitted to ACM TOMS}, 2015.

\bibitem[Robins et~al.(2003)Robins, Rountree, and Rountree]{robins2003learning}
A.~Robins, J.~Rountree, and N.~Rountree.
\newblock {Learning and teaching programming: A review and discussion}.
\newblock \emph{Computer Science Education}, 13\penalty0 (2):\penalty0
  137--172, 2003.

\bibitem[Stefik and Siebert(2013)]{StefikSiebert_2013}
A.~Stefik and S.~Siebert.
\newblock {An Empirical Investigation into Programming Language Syntax}.
\newblock \emph{{ACM Transactions on Computing Education (TOCE)}}, 13\penalty0
  (4), 2013.
\newblock \doi{10.1145/2534973}.

\bibitem[{TIOBE Software}(2015)]{TIOBE_2015}
{TIOBE Software}.
\newblock {TIOBE Programming Community Index:
  www.tiobe.com/index.php/tiobe\_index (last accessed 16 May 2015)}, 2015.

\bibitem[{van Rossum} et~al.(2001){van Rossum}, Warsaw, and Coghlan]{PEP8_2001}
G.~{van Rossum}, B.~Warsaw, and N.~Coghlan.
\newblock {PEP 0008 -- Style Guide for Python Code:
  https://www.python.org/dev/peps/pep-0008/ (last accessed: 16 May 2015)},
  2001.

\bibitem[Wilson(2006)]{Wilson_2006}
G.~Wilson.
\newblock {Software Carpentry: Getting Scientists to Write Better Code by
  Making Them More Productive}.
\newblock \emph{Computing in Science \& Engineering}, November--December 2006.

\bibitem[Wilson(2014)]{Wilson_2014}
G.~Wilson.
\newblock {Software Carpentry: lessons learned}.
\newblock \emph{F1000Research}, 3\penalty0 (62), 2014.
\newblock \doi{10.12688/f1000research.3-62.v1}.

\bibitem[Winslow(1996)]{winslow1996programming}
L.~E. Winslow.
\newblock {Programming pedagogy --- a psychological overview}.
\newblock \emph{ACM SIGCSE Bulletin}, 28\penalty0 (3):\penalty0 17--22, 1996.

\bibitem[Yapici and Akbayin(2012)]{YapiciAkbayin_2012}
I.Umit Yapici and Hasan Akbayin.
\newblock {The Effect of Blended Learning Model on High School Students'
  Biology Achievement and on Their Attitudes towards the Internet}.
\newblock \emph{Turkish Online Journal of Educational Technology}, 11\penalty0
  (2):\penalty0 228--237, 2012.

\bibitem[Yuan and Powell(2013)]{YuanPowell_2013}
L.~Yuan and S.~Powell.
\newblock {MOOCs and Open Education: Implications for Higher Education}.
\newblock \emph{{Centre for Educational Technology, Interoperability and
  Standards}}, pages 7--8, 2013.

\end{thebibliography}

\section*{Appendix A: SOLE lecturer and module scoring criteria}\label{sect:appendix}

Responses to the criteria below were requested as part of SOLE with respect to both the module and the lecturer. Note that in 2010, 2011 and 2012, the available responses were ``Very Good, Good, Satisfactory, Poor, Very Poor, No response'', whereas in 2013 and 2014 the available responses were ``Definitely Agree, Mostly Agree, Neither Agree or Disagree, Mostly Disagree, Definitely Disagree, Not applicable''. In addition to the criteria, text boxes were also provided such that students could provide additional constructive feedback.

\subsection*{A1. Lecturer score criteria (2010, 2011)}
\begin{enumerate}
   \item The structure and delivery of the lectures
   \item The explanation of concepts given by the lecturer
   \item The approachability of the lecturer
   \item The interest and enthusiasm generated by the lecturer
\end{enumerate}

\subsection*{A2. Module score criteria (2010, 2011)}
\begin{enumerate}
   \item The support materials available for this module (e.g. handouts, blackboard/web pages, problem sheets and/or notes on the board)
   \item The organisation of the module
   \item The structure and delivery of the lectures
   \item The explanation of concepts given by the lecturer
   \item The approachability of the lecturer
   \item The interest and enthusiasm generated by the lecturer
\end{enumerate}

\subsection*{A3. Lecturer score criteria (2012)}
\begin{enumerate}
   \item The structure and delivery of the teaching sessions
\end{enumerate}

\subsection*{A4. Module score criteria (2012)}
\begin{enumerate}
   \item The structure and delivery of the teaching sessions
   \item The content of this module
\end{enumerate}

\subsection*{A5. Lecturer score criteria (2013, 2014)}
\begin{enumerate}
   \item The lecturer explained the material well
   \item The lecturer generated interest and enthusiasm
   \item The lecturer was approachable
   \item Overall, I am satisfied with this lecturer
\end{enumerate}

\subsection*{A6. Module score criteria (2013, 2014)}
\begin{enumerate}
   \item The content of the module is well structured
   \item The content of the module is intellectually stimulating 
   \item I have received helpful feedback on my work
   \item Overall, I am satisfied with the quality of the module
\end{enumerate}

\afterpage{%
   \begin{table}
   \caption{Total number of students registered on the programming course each year. In 2010 the programming course was optional and the 35 students represent only a fraction of the total number of first year students in the department that year. In 2012 two separate classes were run: one comprising 73 first year students and one comprising 89 second year students, in order to transition the course from being a second year course to a first year course in later years, hence the larger total number of students.}
   \begin{center}
   \begin{tabular}[t]{r r}
   \hline
   Year & Number of Students\\
   \hline
2010 & 35\\
2011 & 89\\
2012 & 162\\
2013 & 85\\
2014 & 87\\
   \hline
   \end{tabular}
   \end{center}
   \label{table:total_number_of_students}
   \end{table}
}

\afterpage{%
   \begin{table}
   \caption{Total number of A-Level (Further Education) qualifications attained (by subject) by each year's departmental student intake. The 2010 and 2011 programming classes were for second year students, respectively corresponding to the 2009 and 2010 intake. Not all of the students from the 2009 intake took to the programming course in 2010, since it was optional just for that year, which is why the number of A-Levels in 2009 sometimes exceeds the total number of 2010 programming students in Table \ref{table:total_number_of_students}. The 2012 programming class combined both the first and second year students, corresponding to the 2011 and 2012 intake respectively.}
   \begin{center}
   \begin{tabular}[t]{r r r r r r r}
   \hline
   Subject & 2009 & 2010 & 2011 & 2012 & 2013 & 2014\\
   \hline
Maths & 51 & 68 & 71 & 67 & 75 & 74\\
Physics & 37 & 53 & 57 & 49 & 55 & 60\\
Chemistry & 38 & 45 & 52 & 51 & 63 & 52\\
Geology & 18 & 24 & 24 & 13 & 35 & 26\\
Biology & 19 & 27 & 25 & 20 & 26 & 22\\
Computing & 2 & 0 & 1 & 0 & 2 & 0\\
Other & 46 & 79 & 39 & 70 & 74 & 16\\
   \hline
   \end{tabular}
   \end{center}
   \label{table:alevels}
   \end{table}
}

\afterpage{%
   \begin{table}
   \caption{Demographic information for each year's departmental student intake. Note that `EU' (European Union) does not include the UK here, and `Overseas' denotes any non-EU or non-UK country, except qualifying overseas territories determined by The Education (Fees and Awards) (England) Regulations 2007. The 2010 and 2011 programming classes were for second year students, respectively corresponding to the 2009 and 2010 intake. The 2012 programming class combined both the first and second year students, corresponding to the 2011 and 2012 intake respectively.}
   \begin{center}
   \begin{tabular}[t]{r r r}
   \hline
   Year & Gender (male : female) & Student status (UK : EU : Overseas)\\
   \hline
2009 & 39:25 & 55:0:9\\
2010 & 56:33 & 79:3:7\\
2011 & 50:39 & 68:10:11\\
2012 & 44:29 & 61:3:9\\
2013 & 53:32 & 64:4:17\\
2014 & 53:34 & 60:5:22\\
   \hline
   \end{tabular}
   \end{center}
   \label{table:demographics}
   \end{table}
}

\afterpage{%
   \begin{table}
   \caption{Total number of Graduate Teaching Assistants (GTAs) who helped in the programming course each year. In 2013 and 2014 there were a larger total number of GTAs assisting. In these years, the GTAs took turns and operated a rota system, to ensure that approximately 8--10 GTAs were present in each workshop.}
   \begin{center}
   \begin{tabular}[t]{r r}
   \hline
   Year & Number of GTAs\\
   \hline
2010 & 4\\
2011 & 7\\
2012 & 9\\
2013 & 18\\
2014 & 14\\
   \hline
   \end{tabular}
   \end{center}
   \label{table:gtas}
   \end{table}
}

\afterpage{%
   \begin{table}
   \caption{Module (left) and lecturer (right) scores for various criterion, in the year 2010. An average score of -2 to -1.5 indicates `Very Poor', -1.5 to -0.5 indicates `Poor', -0.5 to 0.5 indicates `Satisfactory', 0.5 to 1.5 indicates `Good' and 1.5 to 2 indicates `Very Good'. The relatively high scores suggest that the traditional lecturing style met the students' expectations here.}
   \begin{minipage}[t]{0.48\textwidth}
   \begin{center}
   \begin{tabular}[t]{r r}
   \hline
   Criterion & Score\\
   \hline
   
Support material & 1.20\\
Organisation & 0.89\\
Structure/delivery & 1.11\\
Explanation & 1.11\\
Approachability & 1.65\\
Interest generated & 1.26\\
Mean & 1.20 (Good)\\

   \hline
   \end{tabular}
   \end{center}
   \end{minipage}
   \begin{minipage}[t]{0.48\textwidth}
   \begin{center}
   \begin{tabular}[t]{r r}
   \hline
   Criterion & Score\\
   \hline
   
Structure/delivery & 1.11\\
Explanation & 1.11\\
Approachability & 1.65\\
Interest generated & 1.26\\
Mean & 1.28 (Good)\\

   \hline
   \end{tabular}
   \end{center}
   \end{minipage}
   \label{table:scores_2010}
   \end{table}

   \begin{table}
   \caption{Module (left) and lecturer (right) scores for various criterion, in the year 2011. An average score of -2 to -1.5 indicates `Very Poor', -1.5 to -0.5 indicates `Poor', -0.5 to 0.5 indicates `Satisfactory', 0.5 to 1.5 indicates `Good' and 1.5 to 2 indicates `Very Good'. The relatively high scores suggest that the traditional lecturing style met the students' expectations here.}
   \begin{minipage}[t]{0.48\textwidth}
   \begin{center}
   \begin{tabular}[t]{r r}
   \hline
   Criterion & Score\\
   \hline
   
Support material & 0.58\\
Organisation & 0.82\\
Structure/delivery & 0.65\\
Explanation & 0.37\\
Approachability & 1.10\\
Interest generated & 0.95\\
Mean & 0.75 (Good)\\

   \hline
   \end{tabular}
   \end{center}
   \end{minipage}
   \begin{minipage}[t]{0.48\textwidth}
   \begin{center}
   \begin{tabular}[t]{r r}
   \hline
   Criterion & Score\\
   \hline
   
Structure/delivery & 0.65\\
Explanation & 0.37\\
Approachability & 1.10\\
Interest generated & 0.95\\
Mean & 0.77 (Good)\\

   \hline
   \end{tabular}
   \end{center}
   \end{minipage}
   \label{table:scores_2011}
   \end{table}

   \begin{table}
   \caption{Module (left) and lecturer (right) scores for various criterion, in the year 2012. An average score of -2 to -1.5 indicates `Very Poor', -1.5 to -0.5 indicates `Poor', -0.5 to 0.5 indicates `Satisfactory', 0.5 to 1.5 indicates `Good' and 1.5 to 2 indicates `Very Good'. The much lower scores here suggest that the students were put off by the flipped classroom format since it was not what they were used to.}
   \begin{minipage}[t]{0.48\textwidth}
   \begin{center}
   \begin{tabular}[t]{r r}
   \hline
   Criterion & Score\\
   \hline
   
Structure/delivery & 0.38\\
Content & 0.49\\
Mean & 0.44 (Satisfactory)\\

   \hline
   \end{tabular}
   \end{center}
   \end{minipage}
   \begin{minipage}[t]{0.48\textwidth}
   \begin{center}
   \begin{tabular}[t]{r r}
   \hline
   Criterion & Score\\
   \hline
   
Structure/delivery & 0.38\\
Mean & 0.38 (Satisfactory)\\

   \hline
   \end{tabular}
   \end{center}
   \end{minipage}
   \label{table:scores_2012}
   \end{table}

   \begin{table}
   \caption{Module (left) and lecturer (right) scores for various criterion, in the year 2013. An average score of -2 to -1.5 indicates `Very Poor', -1.5 to -0.5 indicates `Poor', -0.5 to 0.5 indicates `Satisfactory', 0.5 to 1.5 indicates `Good' and 1.5 to 2 indicates `Very Good'. The much lower scores here (particularly for the ``Explanation'' criterion) suggest that the students were put off by the flipped classroom format since it was not what they were used to.}
   \begin{minipage}[t]{0.48\textwidth}
   \begin{center}
   \begin{tabular}[t]{r r}
   \hline
   Criterion & Score\\
   \hline
   
Structure & 0.45\\
Intellectual stimulation & 0.72\\
Feedback & 0.47\\
Overall satisfaction & 0.36\\
Mean & 0.50 (Satisfactory)\\

   \hline
   \end{tabular}
   \end{center}
   \end{minipage}
   \begin{minipage}[t]{0.48\textwidth}
   \begin{center}
   \begin{tabular}[t]{r r}
   \hline
   Criterion & Score\\
   \hline
   
Explanation & -0.06\\
Interest generated & 0.67\\
Approachability & 1.33\\
Overall satisfaction & 0.71\\
Mean & 0.66 (Good)\\

   \hline
   \end{tabular}
   \end{center}
   \end{minipage}
   \label{table:scores_2013}
   \end{table}

   \begin{table}
   \caption{Module (left) and lecturer (right) scores for various criterion, in the year 2014. An average score of -2 to -1.5 indicates `Very Poor', -1.5 to -0.5 indicates `Poor', -0.5 to 0.5 indicates `Satisfactory', 0.5 to 1.5 indicates `Good' and 1.5 to 2 indicates `Very Good'. The scores were much higher for this year after reassuring and justifying the blended learning approach to the students.}
   \begin{minipage}[t]{0.48\textwidth}
   \begin{center}
   \begin{tabular}[t]{r r}
   \hline
   Criterion & Score\\
   \hline
   
Structure & 1.26\\
Intellectual stimulation & 1.30\\
Feedback & 1.32\\
Overall satisfaction & 1.21\\
Mean & 1.27 (Good)\\

   \hline
   \end{tabular}
   \end{center}
   \end{minipage}
   \begin{minipage}[t]{0.48\textwidth}
   \begin{center}
   \begin{tabular}[t]{r r}
   \hline
   Criterion & Score\\
   \hline
   
Explanation & 1.11\\
Interest generated & 1.59\\
Approachability & 1.73\\
Overall satisfaction & 1.69\\
Mean & 1.53 (Very good)\\

   \hline
   \end{tabular}
   \end{center}
   \end{minipage}
   \label{table:scores_2014}
   \end{table}

\clearpage
}

\afterpage{%
\begin{figure}
\vspace*{-20mm}
\centering\includegraphics[width=\columnwidth]{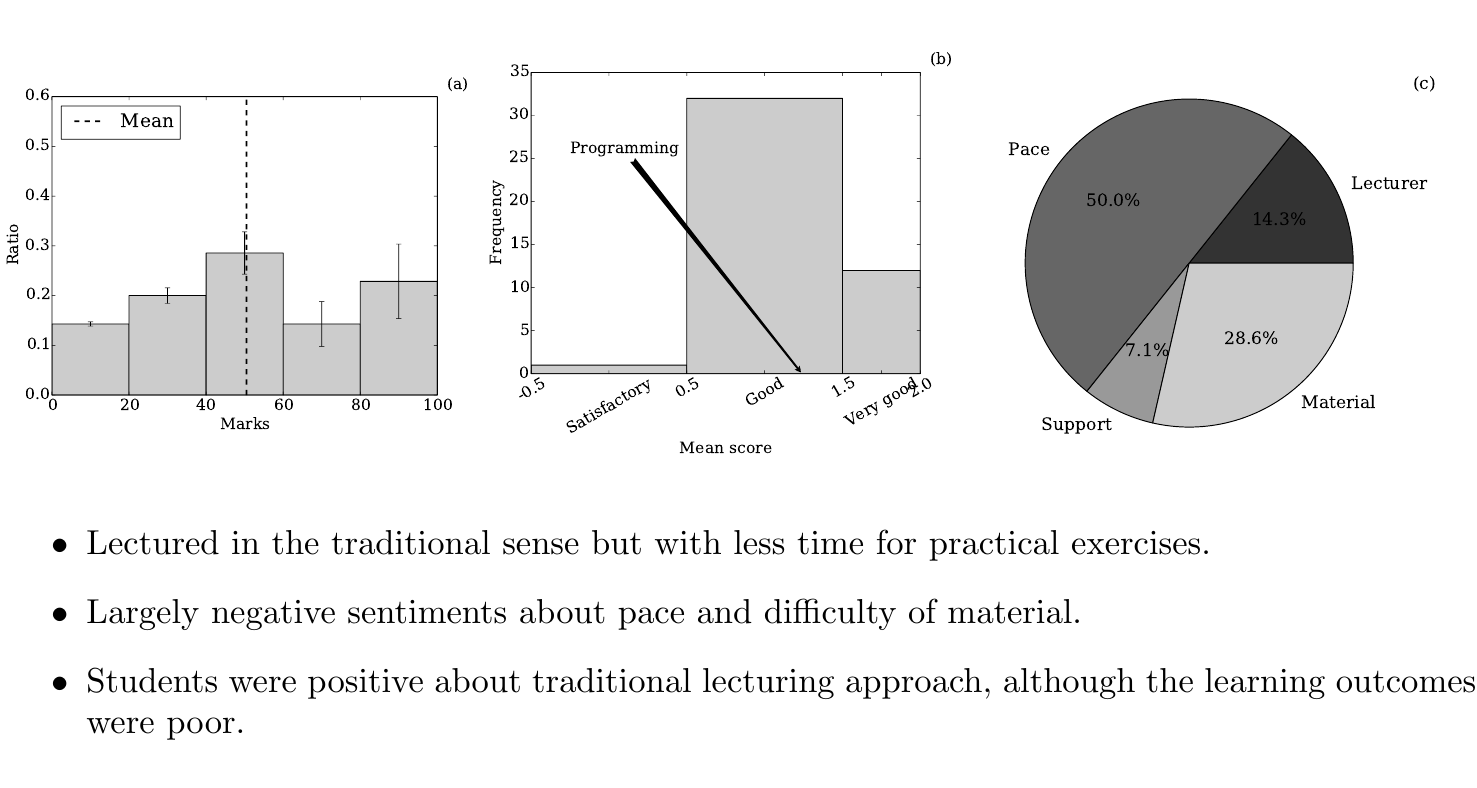}
  \caption{Data for the 2010 class: (a) histogram of final exam marks; (b) histogram showing mean (combined) module and lecturer SOLE scores for all modules in the Department of Earth Science and Engineering; (c) pie chart showing the emergent topics from the SOLE feedback; (bottom row) overall student sentiments. There were 35 students who sat the final exam in 2010. Note that in all years, there were no mean combined (lecturer and module) SOLE scores below `Satisfactory'.}
  \label{fig:data_2010}
\end{figure}

\begin{figure}
\vspace*{-20mm}
\centering\includegraphics[width=\columnwidth]{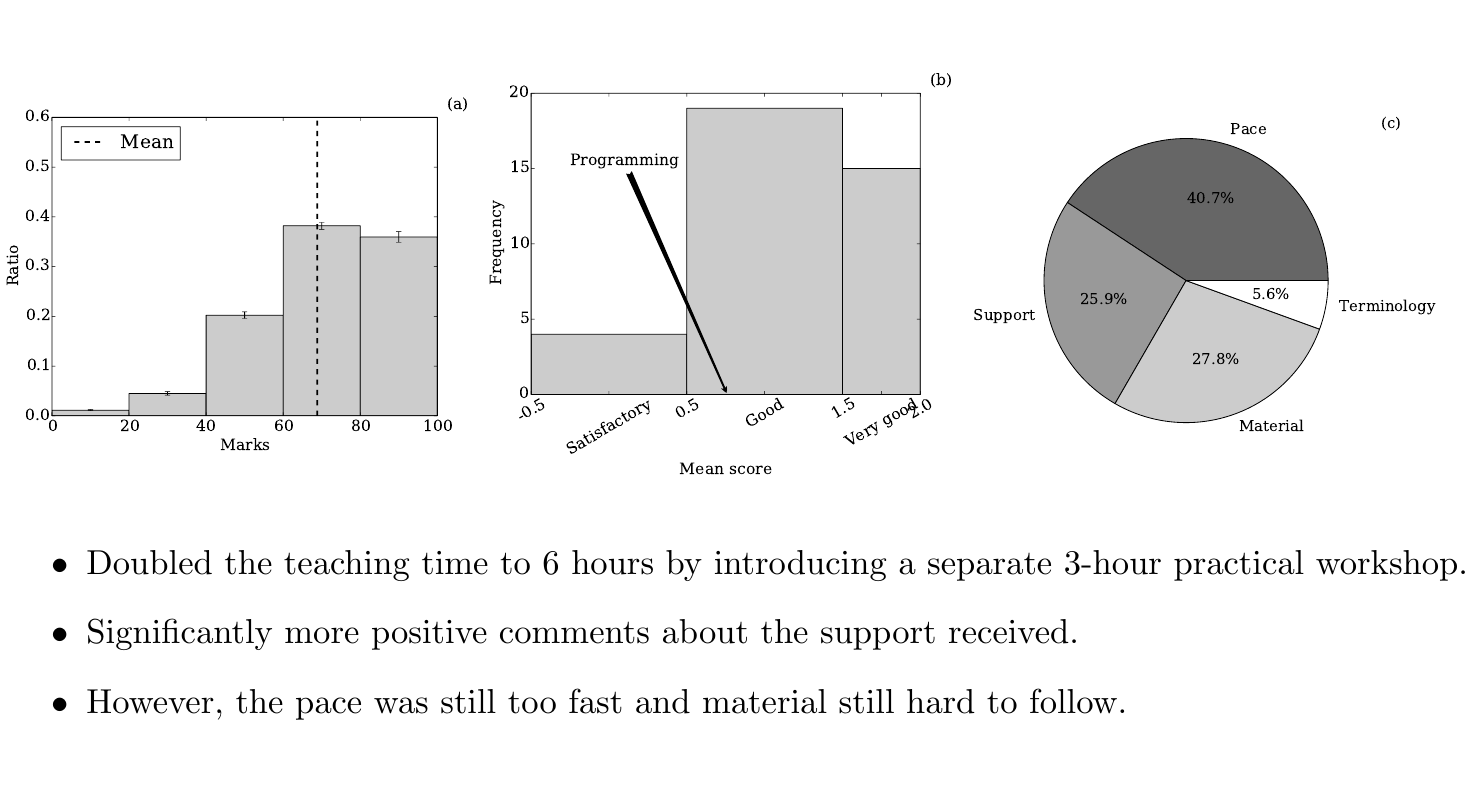}
  \caption{Data for the 2011 class: (a) histogram of final exam marks; (b) histogram showing mean (combined) module and lecturer SOLE scores for all modules in the Department of Earth Science and Engineering; (c) pie chart showing the emergent topics from the SOLE feedback; (bottom row) overall student sentiments. There were 89 students who sat the final exam in 2011. Note that in all years, there were no mean combined (lecturer and module) SOLE scores below `Satisfactory'.}
  \label{fig:data_2011}
\end{figure}

\begin{figure}
\vspace*{-20mm}
\centering\includegraphics[width=\columnwidth]{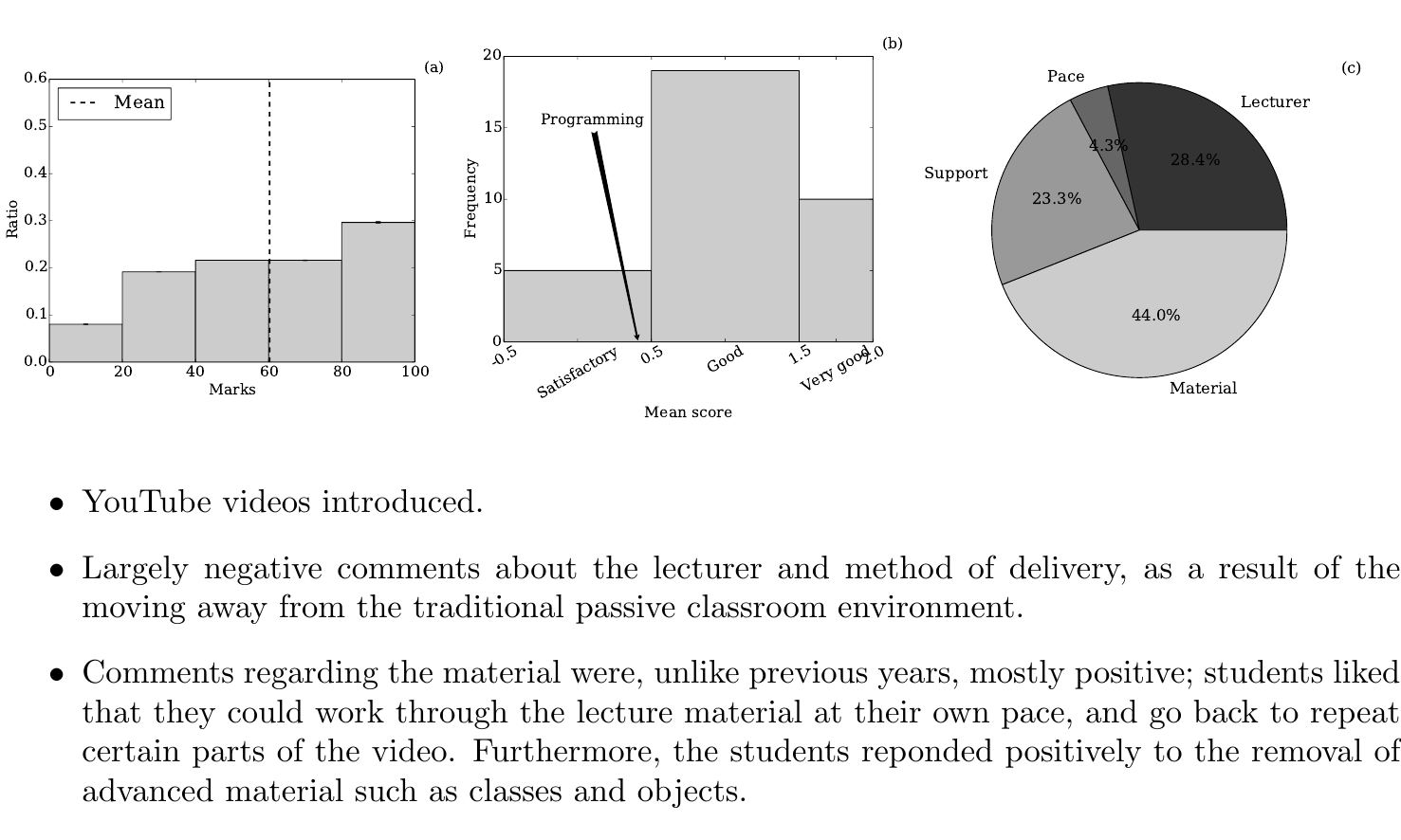}
  \caption{Data for the 2012 class: (a) histogram of final exam marks; (b) histogram showing mean (combined) module and lecturer SOLE scores for all modules in the Department of Earth Science and Engineering; (c) pie chart showing the emergent topics from the SOLE feedback; (bottom row) overall student sentiments. There were 162 (73 first year and 89 second year) students who sat the final exam in 2012. Note that in all years, there were no mean combined (lecturer and module) SOLE scores below `Satisfactory'.}
  \label{fig:data_2012}
\end{figure}

\begin{figure}
\vspace*{-20mm}
\centering\includegraphics[width=\columnwidth]{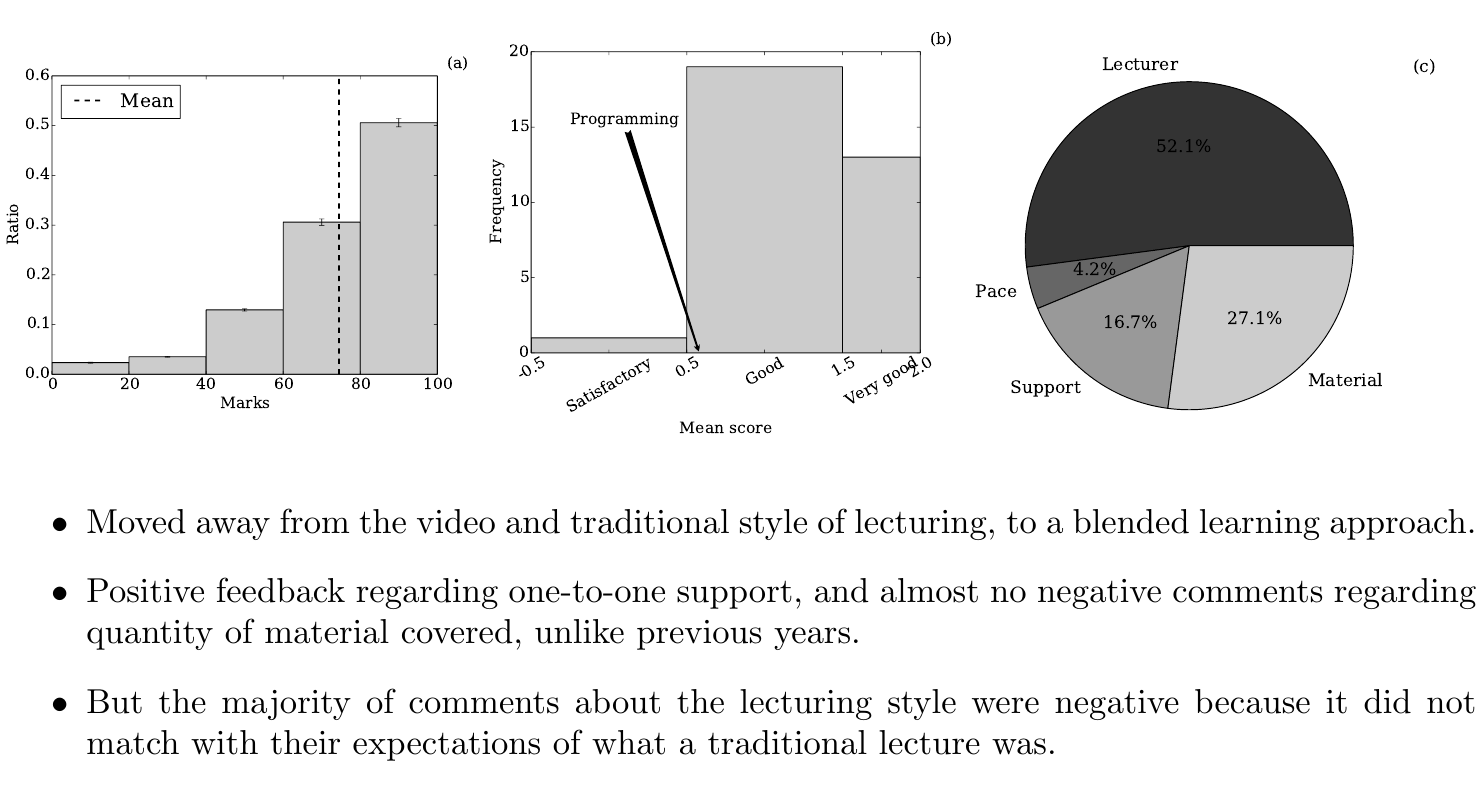}
  \caption{Data for the 2013 class: (a) histogram of final exam marks; (b) histogram showing mean (combined) module and lecturer SOLE scores for all modules in the Department of Earth Science and Engineering; (c) pie chart showing the emergent topics from the SOLE feedback; (bottom row) overall student sentiments. There were 85 students who sat the final exam in 2013. Note that in all years, there were no mean combined (lecturer and module) SOLE scores below `Satisfactory'.}
  \label{fig:data_2013}
\end{figure}

\begin{figure}
\vspace*{-20mm}
\centering\includegraphics[width=\columnwidth]{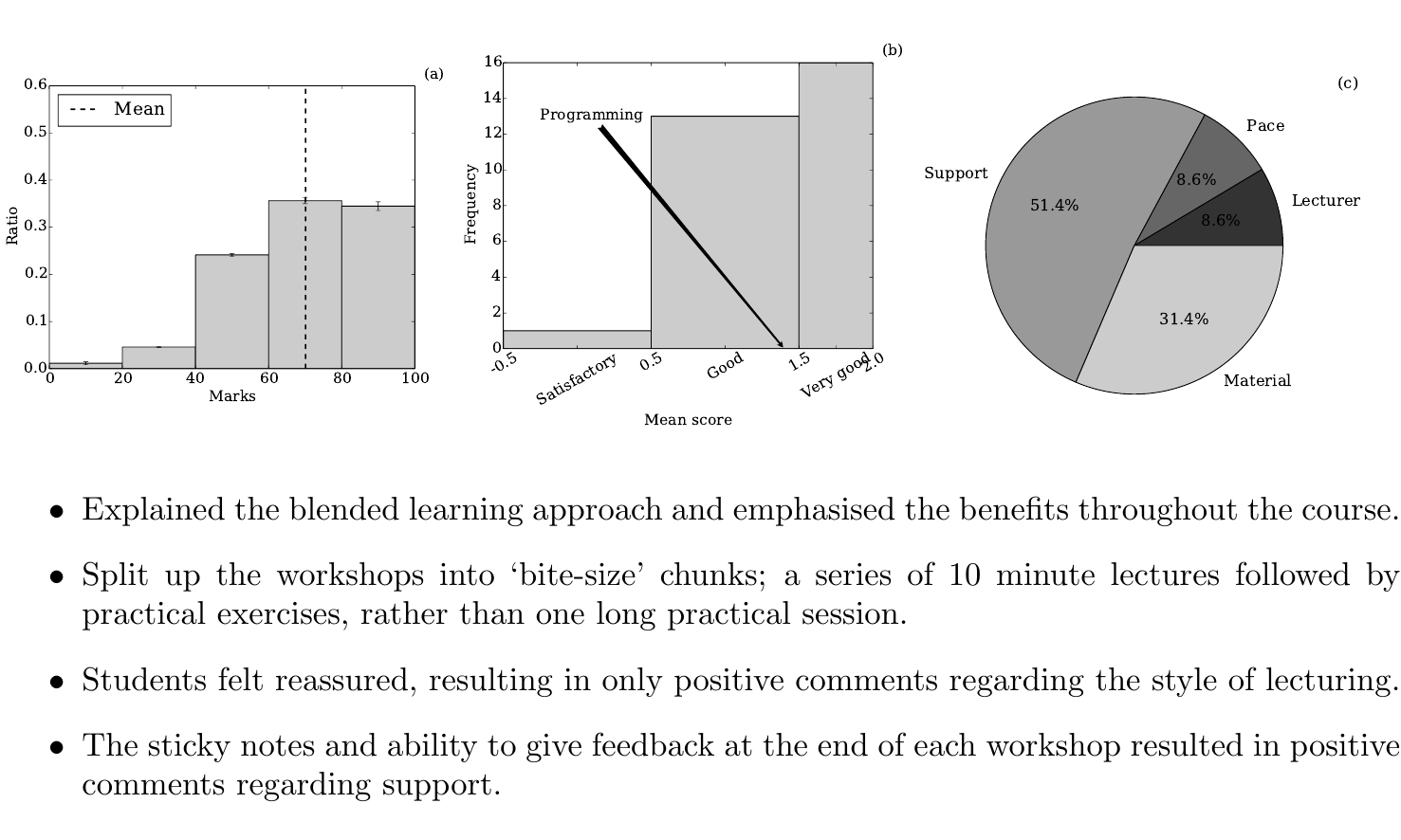}
  \caption{Data for the 2014 class: (a) histogram of final exam marks; (b) histogram showing mean (combined) module and lecturer SOLE scores for all modules in the Department of Earth Science and Engineering; (c) pie chart showing the emergent topics from the SOLE feedback; (bottom row) overall student sentiments. There were 87 students who sat the final exam in 2014. Note that in all years, there were no mean combined (lecturer and module) SOLE scores below `Satisfactory'.}
  \label{fig:data_2014}
\end{figure}

\clearpage
}
\end{document}